\documentstyle[12pt]{article}
\title{\bf String Representation of the\\
Abelian Higgs Model with an\\
Axionic Interaction}  
\author{D.V. ANTONOV \thanks{Phone: 0049-30-2093 7974; Fax: 0049-30-2093 
7631; E-mail address: 
antonov@pha2.physik.hu-berlin.de}~ \thanks {On leave of absence from the 
Institute of Theoretical and Experimental Physics, B.Cheremushkinskaya 25, 
117 218, Moscow, Russia.} 
\\
{\it Institut f\"ur Physik, Humboldt-Universit\"at zu Berlin,}\\
{\it Invalidenstrasse 110, D-10115, Berlin, Germany}}
\date{}
\begin{document}
\maketitle
\vspace{1mm}
\centerline{\bf {Abstract}}
\vspace{3mm}
Making use of the duality transformation, we construct string 
representation for the partition function of the 
London limit of Abelian Higgs Model  
with an additional axionic term. 
In the lowest order of perturbation theory, 
this term leads to the appearance in the resulting string 
effective action 
of a new threelocal interaction between the  
elements of the string world-sheet. Consequently, there emerges 
a threelocal correlator of the dual field strength tensors, which does 
not contain the average over world-sheets, and is therefore 
nontrivial even in the sector of the theory with a single small 
vortex. The relation between the obtained correlator and the 
bilocal one is established. Finally, it is argued that the vacuum 
structure of the London 
limit of the Abelian Higgs Model with an additional axionic interaction  
is much more 
similar to that of gluodynamics rather than without this interaction.

\vspace{3mm}

\vspace{3mm}

\newpage

In the last years, there appeared a lot of papers concerning  
string 
representation of the Abelian Higgs Model (AHM) [1-4]. 
Such an interest to this theory  
is motivated by the original conjecture of 't Hooft and 
Mandelstam [5] that on the phenomenological level, quark confinement 
in QCD could be with a good accuracy described as a dual Meissner effect 
in superconductor, and AHM is just the 4D model of the latter. 

The main 
analytical approach to the problem under study is the so-called duality 
transformation [6]. This transformation enables one to 
reformulate an integral over 
the regular part of the phase of the Higgs field as an integral over 
an antisymmetric tensor field, which eventually acquires a mass equal 
to the mass of the gauge boson generated by the Higgs mechanism 
\footnote{
It is worth noting, that 
analogous partition functions of the massive antisymmetric tensor field 
interacting with the string have been for the first time proposed already 
in the first paper of Ref. [1] and in Ref. [7] for investigation 
of a local theory of charges 
and monopoles. They have also been considered in Ref. [8]  
for the purposes of construction of 
a superconducting medium with an antisymmetric 
tensor order parameter. The same theory 
recently also 
occured to be relevant to the dual description of 3D compact 
QED [9].}. On 
the other hand, an integral over the singular part of the phase 
of the Higgs field 
can be rewritten as an integral over the surfaces, at 
which the Higgs field is singular. These surfaces are nothing else, 
but the world-sheets of closed vortices, which are present in the 
type-II superconductor and consequently in the London limit of AHM. 
Such vortices are usually 
called 
Abrikosov-Nielsen-Olesen 
(ANO) strings. 
In this way, we get 
an interaction 
of the massive antisymmetric tensor field with the string world-sheet, 
which represents 
a coupling of the gauge boson to the ANO string. 
Integrating over this tensor field, one eventually arrives at 
the action of AHM in the London limit in the form of the  
interaction of 
{\it two} elements of the ANO string world-sheet via an exchange of 
the gauge boson. The resulting partition function reads 

$${\cal Z}_{\rm AHM}=\int {\cal D}x_\mu (\xi)\exp\left\{-\int 
\limits_\Sigma^{} d\sigma_{\mu\nu} 
(x(\xi))\int\limits_\Sigma^{} 
d\sigma_{\lambda\rho} (x(\xi'))D_{\mu\nu, \lambda\rho} 
(x(\xi)-x(\xi'))\right\}. $$
Here $\Sigma$ is a world-sheet of the ANO string parametrized by 
$x_\mu(\xi)$, $\xi=\left(\xi^1, \xi^2\right)$, and 
$D_{\mu\nu, \lambda\rho} 
(x(\xi)-x(\xi'))$ stands for the propagator of the massive antisymmetric 
field (the so-called Kalb-Ramond field). From now on, we for simplicity 
omit the Jacobian emerging during the change 
of the integration variables $\theta^{{\rm sing.}}\to x_\mu(\xi)$, which 
for the case when $\Sigma$ has a spherical topology has been calculated 
in Ref. [2]. 

The aim of the present Letter is to derive string representation for  
the partition function of AHM with an additional term which describes 
an interaction of an axion with two photons. This term is in fact 
a $\theta$-term, where $\theta$ however is not a $c$-number 
parameter as it took place in Ref. [3], 
but a phase of the Higgs field. Such a term emerges due to the 
integration over the heavy fermions interacting with the phase 
$\theta$ in the following manner [10] 

$${\cal L}_{\rm int.}=\eta\bar\psi\left(\cos\theta+i\gamma_5\sin\theta
\right)\psi,$$
where $\eta$ stands for the square root of the v.e.v. of the Higgs 
field. 
Notice, that 
some remarks concerning dual formulation of the theory 
of a complex scalar field initially 
possessing a {\it global} $U(1)$-symmetry, which is then broken by    
the axionic interaction, have been presented already in Ref. [6].    
In what follows, we shall derive a dual formulation of 
AHM (which is a theory with a {\it local} $U(1)$-symmetry) 
in the London limit with the axionic term added. In particular, it will 
be demonstrated that in the lowest order of 
perturbation theory this term leads 
to the appearance in the resulting string effective 
action of the interaction of {\it three} world-sheet elements. 
This interaction gives rise to a {\it threelocal} correlator of the 
dual field strength tensors, which does not contain the average 
over string world-sheets. Without axionic interaction, 
such a correlator has been demonstrated to be absent even in 
vicinity of the London limit [11]. We shall also establish a relation 
between the obtained threelocal correlator and the bilocal one. This 
relation occurs to be 
quite similar to the one which holds in gluodynamics [12], which makes 
it possible to conclude that 
the structure of vacuum of the London limit of AHM 
with an additional axionic term is much more similar to that of 
gluodynamics than without such a term.

Let us start with the partition function of the London limit of 
AHM with an additional 
axionic term (for which we make use of the notations of Ref. [10]). 
This partition function reads 

$${\cal Z}_{\rm AHM}^{\rm axionic}=
\int {\cal D}A_\mu {\cal D}\theta^{\rm sing.} 
{\cal D}\theta^{\rm reg.}\exp\left\{-\int d^4x\left[\frac14 F_{\mu\nu}^2
+\frac{\eta^2}{2}\left(\partial_\mu\theta-eA_\mu\right)^2+
\frac{e^2\theta}{32\pi^2}\varepsilon_{\mu\nu\lambda\rho}F_{\mu\nu}
F_{\lambda\rho}\right]\right\}. \eqno (1)$$
In Eq. (1), the phase of the Higgs field has been decomposed as follows, 
$\theta=\theta^{\rm reg.}+\theta^{\rm sing.}$, where 
$\theta^{\rm reg.}$ describes a single-valued fluctuation around 
the vortex configuration, whereas 
$\theta^{\rm sing.}$ describes a given configuration of vortices. The 
singular part of the phase of the Higgs field is 
unambiguously related to the string world-sheet 
$\Sigma$ via the equation [6] 

$$\varepsilon_{\mu\nu\lambda\rho}\partial_\lambda
\partial_\rho\theta^{\rm sing.}(x)=2\pi\Sigma_{\mu\nu}(x), \eqno (2)$$
where 
$\Sigma_{\mu\nu}(x)\equiv\int\limits_\Sigma^{}d\sigma_{\mu\nu}(x(\xi))
\delta (x-x(\xi))$ is the vorticity 
tensor current defined on $\Sigma$, which is 
conserved, i.e. 
$\partial_\mu\Sigma_{\mu\nu}=0$, since $\Sigma$ is a closed 
surface.

Making use of Eq. (2) and performing the same 
duality transformation as the one which has been proposed 
in Ref. [6] and applied in Refs. [1-4], we get from Eq. (1) 
the following expression for the partition function 
 
$${\cal Z}_{\rm AHM}^{\rm axionic}=
\int {\cal D}A_\mu {\cal D}h_{\mu\nu}{\cal D}x_\mu (\xi)
\exp\left\{\int d^4x\left[-\frac{1}{12\eta^2}H_{\mu\nu\lambda}^2
+i\pi h_{\mu\nu}\Sigma_{\mu\nu}-\right.\right.$$

$$\left.\left.-\frac14 F_{\mu\nu}^2
-\frac{ie}{2}\varepsilon_{\mu\nu\lambda\rho}A_\mu\partial_\nu 
h_{\lambda\rho}
-\frac{ie^2}{8\pi^2\eta^2}H_{\mu\nu\lambda}A_\mu\partial_\nu A_\lambda
+{\cal O}\left(e^4\right)\right]\right\}. \eqno (3)$$
In Eq. (3), $h_{\mu\nu}$ stands for an antisymmetric tensor field 
dual to $\theta^{\rm reg.}$, $H_{\mu\nu\lambda}=\partial_\mu 
h_{\nu\lambda}+\partial_\lambda h_{\mu\nu}+\partial_\nu h_{\lambda\mu}$ 
is its strength tensor, and the integration over the string world-sheets 
emerged from the integration over $\theta^{\rm sing.}$ in a sense 
described above.  
All the terms in the argument of the exponent standing on the R.H.S. 
of Eq. (3) except for the one, which is of the order 
of $e^2$ are the result 
of the duality transformation of the partition function of AHM without 
the axionic interaction [1-4], whereas the term of the order of $e^2$, 
which has emerged 
just due to this interaction, is a new one. Notice also, that 
since from now on   
we shall be interested only in the terms following from the axionic 
interaction 
in the lowest order of perturbation theory, on the R.H.S. 
of Eq. (3) we have omitted the terms of the order of $e^4$.

It is now technically convenient to pass from the integration 
over $h_{\mu\nu}$ to 
the integration over the field $\varphi_\mu\equiv
\varepsilon_{\mu\nu\lambda\rho}\partial_\nu h_{\lambda\rho}$. Since 
such a change of variables is linear, the resulting 
Jacobian factor will be trivial (i.e. it will not affect string 
effective action, we are looking for), and therefore in what follows we 
shall not be interested in it. In terms of the new field, the partition 
function reads as follows 

$${\cal Z}_{\rm AHM}^{\rm axionic}=
\int {\cal D}A_\mu {\cal D}\varphi_\mu {\cal D}x_\mu (\xi)
\exp\left\{\int d^4x\left[-\frac{1}{8\eta^2}\varphi_\mu^2+
\frac{i}{3\pi}\varepsilon_{\mu\nu\lambda\rho}\Sigma_{\mu\nu}(x)
\int d^4y\frac{(x-y)_\lambda}{|x-y|^4}\varphi_\rho (y)-\right.\right.$$

$$\left.\left.-\frac14 F_{\mu\nu}^2-\frac{ie}{2}\varphi_\mu A_\mu+
\frac{ie^2}{16\pi^2\eta^2}\varepsilon_{\mu\nu\lambda\rho}
\varphi_\rho A_\mu\partial_\nu A_\lambda
+{\cal O}\left(e^4\right)\right]\right\}. \eqno (4)$$

Now, carrying out the integration over the field $\varphi_\mu$, we 
get from Eq. (4) 

$${\cal Z}_{\rm AHM}^{\rm axionic}=
\int {\cal D}x_\mu(\xi)\exp\left[-\frac{2\eta^2}{9}
\int\limits_\Sigma^{}d\sigma_{\mu\nu}(x) 
\int\limits_\Sigma^{}d\sigma_{\mu\nu}(y)\frac{1}{(x-y)^2}\right]
\cdot$$

$$\cdot\int {\cal D}A_\mu\exp\left\{-\int d^4x\left[\frac14 F_{\mu\nu}^2
+\frac{m^2}{2}A_\mu^2+\frac{2m\eta}{3\pi}\varepsilon_{\mu\nu\lambda
\rho}\Sigma_{\mu\nu}(x)\int d^4y\frac{(y-x)_\lambda}{|y-x|^4}
A_\rho(y)+\right.\right.$$

$$\left.\left.+\frac{e^2}{6\pi^3}\left(A_\mu\partial_\nu A_\lambda+
A_\lambda\partial_\mu A_\nu+A_\nu\partial_\lambda A_\mu\right)
\int d^4y \frac{(y-x)_\lambda}{|y-x|^4}\Sigma_{\mu\nu}(y)+
{\cal O}\left(e^4\right)\right]\right\}, \eqno (5)$$
where $m\equiv e\eta$ is the mass of the gauge boson generated by 
the Higgs mechanism. Finally, integration over the field $A_\mu$ in 
Eq. (5) yields the desirable expression for the partition function 

$${\cal Z}_{\rm AHM}^{\rm axionic}=
\int {\cal D}x_\mu(\xi)\exp\left\{-\frac{2\eta^2}{9}
\int\limits_\Sigma^{}d\sigma_{\mu\nu}(x) 
\int\limits_\Sigma^{}d\sigma_{\mu\nu}(y)\frac{1}{(x-y)^2}-\right.$$
 
$$-\frac{m^3\eta^2}{9\pi^4}\int\limits_\Sigma^{} d\sigma_{\mu\nu}
(z)\int\limits_\Sigma^{}
d\sigma_{\mu\lambda}(u)\int d^4xd^4y\frac{K_1\left(m|x-y|\right)}
{|x-y|}\left[2\frac{(x-z)_\lambda (y-u)_\nu}{|x-z|^4 |y-u|^4}-
\delta_{\nu\lambda}\frac{(x-z)_\rho (y-u)_\rho}{|x-z|^4 |y-u|^4}
\right]-$$

$$-\frac{m^6}{54\pi^9}\int\limits_\Sigma^{}d\sigma_{\mu\nu}(x)\int 
d^4y\frac{(y-x)_\lambda}{|y-x|^4}\int d^4wd^4u
\Biggl[\int\limits_\Sigma^{} 
d\tilde\sigma_{\rho\nu}
(v)\int\limits_\Sigma^{}
d\tilde\sigma_{\sigma\mu}(z)\frac{(w-v)_\rho}{|w-v|^4}
\frac{(u-z)_\sigma}{|u-z|^4}\frac{(y-u)_\lambda}{(y-u)^2}+$$

$$+(\nu\to\mu, \lambda\to\nu, \mu\to\lambda)+(\nu\to\lambda, 
\lambda\to\mu, \mu\to\nu)\Biggr]\cdot$$

$$\left.\cdot\frac{K_1\left(m|y-w|\right)}{|y-w|}\left[\frac{K_1\left(
m|y-u|\right)}{|y-u|}+\frac{m}{2}\Biggl(K_0\left(m|y-u|\right)+
K_2\left(m|y-u|\right)\Biggr)\right]+{\cal O}\left(e^8\right)
\right\}, \eqno (6)$$
where $d\tilde\sigma_{\mu\nu}\equiv\frac12\varepsilon_{\mu\nu\lambda\rho}
d\sigma_{\lambda\rho}$, and 
$K_i$'s, $i=0,1,2$, stand for the modified Bessel functions.

The first two terms in the string effective action standing in the 
exponent on the R.H.S. of Eq. (6) are the same as the ones which 
could be obtained in the London limit of AHM without an additional 
axionic term, if one integrates first over the field $\varphi_\mu$ and 
then over the field $A_\mu$. The difference of these terms 
from the string effective action obtained in Refs. [1-4] is due 
to the fact that the string effective action derived in these Refs. 
was obtained in a certain gauge for the field $h_{\mu\nu}$, which 
we have not applied here. 

The third term in the exponent on the R.H.S. of Eq. (6), 
which follows from the axionic part of the 
partition function (1) and contains threelocal interactions 
of the elements of the string world-sheet, 
is a quite new one. 
Such a term has been 
absent previously both in the string representation of the AHM 
partition function in the London limit [2] and beyond this limit, i.e. 
in the $1/\lambda$-expansion [11], where $\lambda$ is the AHM coupling 
constant, since the leading term of this 
expansion is quartic in the string world-sheet elements 
\footnote{Notice, that 
the axionic interaction contributes to the quartic terms as well, and 
they arise already in the order of $e^8$. Therefore, would one study 
AHM in the vicinity of the London limit, there emerge two types of terms 
quartic in the string world-sheet elements.}. In order to understand 
why this threelocal term is important, let us consider the 
generating functional of the dual field strength tensors in the 
London limit. It can be obtained from Eq. (1) by adding to 
the argument of the exponent on its R.H.S. the term $-i\int d^4x 
S_{\mu\nu}\tilde F_{\mu\nu}$, 
where $S_{\mu\nu}$ 
stands for the source of the dual field strength tensor \footnote{
See Ref. [11], where it has been in particular explained, why it is 
necessary to consider correlators of the dual rather than of the 
ordinary field strength tensors.}. 
This results in the 
following shift of the vorticity tensor current in the final Eq. (6), 
$\Sigma_{\mu\nu}\to\Sigma_{\mu\nu}+\frac{ie}{\pi}S_{\mu\nu}$. 
In what follows, we shall similarly to Ref. [11] restrict 
ourselves to the sector of AHM in the London limit with a single 
vortex, whose area $\left|\Sigma\right|$ obeys the  
inequality $e\eta^2\left|\Sigma\right|\ll 1$. For such a vortex, 
all the terms in the string representation for correlators 
of the dual field strength tensors containing the average over world-sheets 
can be disregarded w.r.t. the other ones. In this way, we obtain from 
Eq. (6) the following correction to the generating functional of the 
dual field strength tensors, which gives rise to the leading 
contribution to the threelocal correlator,

$$\Delta {\cal Z}_{\rm AHM}^{\rm axionic}\left[S_{\alpha\beta}\right]=
{\cal Z}_{\rm AHM}^{\rm axionic}\left[0\right]
\exp\left\{\frac{ie^3m^2}{27\pi^8}\int d^4xd^4vd^4z S_{\mu\nu}(x)
S_{\alpha\beta}(v)S_{\gamma\delta}(z)\cdot\right.$$

$$\left.\cdot\int d^4yd^4wd^4u
\frac{(y-x)_\lambda}{\left|y-x\right|^4}\frac{(w-v)_\rho}{\left|w-v
\right|^4}\frac{ 
(u-z)_\sigma}{\left|u-z\right|^4}
\Biggl[
\varepsilon_{\rho\nu\alpha\beta}
\varepsilon_{\sigma\mu\gamma\delta}(y-u)_\lambda+
(\nu\to\mu, \lambda\to\nu, \mu\to\lambda)+\right.$$

$$\left.+(\nu\to\lambda, 
\lambda\to\mu, \mu\to\nu)\Biggr]D\left( (y-w)^2\right)D_1\left(
(y-u)^2\right)\right\}, \eqno (7)$$
where ${\cal Z}_{\rm AHM}^{\rm axionic}\left[0\right]$ is given 
by Eq. (6). In Eq. (7), $D$ and $D_1$ stand for the 
renormalization group-invariant coefficient functions, which 
according to the Stochastic Vacuum Model approach [13] parametrize 
the bilocal correlator of the dual field strength tensors 
as follows  

$$\left<\tilde F_{\lambda\nu}(x)\tilde F_{\mu\rho}(0)\right>=
\Biggl(\delta_{\lambda\mu}\delta_{\nu\rho}-\delta_{\lambda\rho}
\delta_{\nu\mu}\Biggr)D\left(x^2\right)+$$

$$+\frac12\Biggl[\partial_\lambda
\Biggl(x_\mu\delta_{\nu\rho}-x_\rho\delta_{\nu\mu}\Biggr)
+\partial_\nu\Biggl(x_\rho\delta_{\lambda\mu}-x_\mu\delta_{\lambda\rho}
\Biggr)\Biggr]D_1\left(x^2\right).$$
Due to Ref. [11], for the case of the London limit of AHM, these functions 
are equal to

$$D\left(x^2\right)=\frac{m^3}{4\pi^2}\frac{K_1(m|x|)}{\left|x\right|}$$
and 

$$D_1\left(x^2\right)=\frac{m}{2\pi^2x^2}\Biggl[
\frac{K_1(m|x|)}{\left|x\right|}
+\frac{m}{2}\Biggl(K_0(m|x|)+K_2(m|x|)\Biggr)\Biggr].$$ 
Then, the leading contribution to the threelocal correlator can be 
obtained from Eq. (7) according to the formula 

$$\left.\left<\tilde F_{\mu_1\nu_1}(x_1)\tilde F_{\mu_2\nu_2}(x_2)
\tilde F_{\mu_3\nu_3}(x_3)\right>=\frac{1}{(-i)^3{\cal Z_{\rm AHM}^
{\rm axionic}}[0]}
\frac{\delta^3\Delta{\cal Z_{\rm AHM}^{\rm axionic}}\left[S_{\alpha\beta}
\right]}{\delta S_{\mu_1\nu_1}(x_1)\delta S_{\mu_2\nu_2}(x_2) 
\delta S_{\mu_3\nu_3}(x_3)}\right|_{S_{\alpha\beta}=0}. $$
The $3!=6$ terms following from Eq. (7) after its threefold 
variation are obvious, and we shall not present them here 
for shortness. 
Since the resulting contribution to the threelocal correlator 
does not contain the average over string 
world-sheets, it is essential even in the sector of AHM under study, 
where we have only one small vortex. 

The nontrivial result is that
by virtue of Eq. (7) this leading contribution can be {\it completely} 
described in terms of the functions $D$ and $D_1$, which parametrize 
the bilocal correlator. This is an example of equation relating 
correlators of different orders to each other \footnote{
Another  
equations for vacuum correlators obtained by several methods 
in various theories, including QCD, have been derived and 
investigated in Refs. [12], [14], and [15].}. 
An 
analogous relation occurs between the quartic and the bilocal 
correlators in the vicinity of the London of AHM, and has been established 
in Ref. [11]. Let us however point out once more that in the absence 
of the axionic interaction, the threelocal correlator which does 
not contain the average over string world-sheets has been lacking 
even in the vicinity of the London limit.  

Finally, as it has been first mentioned in Ref. [12], in the gluodynamics 
case, 
the presence of the 
nonvanishing threelocal correlator is necessary for the consistency of the 
Stochastic Vacuum Model approach. 
Namely, in this theory, the relation   
between the bilocal and threelocal correlators [14], being resolved 
w.r.t. the bilocal correlator in the case of vanishing 
threelocal one, would yield for the function $D$ only 
a constant solution. This is however known to be quite not the case, 
according to the lattice data concerning the function $D$ [16]. 
This means that the so-called Gaussian approximation in the 
Stochastic Vacuum Model [13], according to which all the 
irreducible correlators higher than the bilocal one can be 
with a good accuracy disregarded, is not self-consistent. Namely, it  
must be extended by postulating the nonvanishing threelocal 
correlator as well as the property of factorization of all the 
higher irreducible 
correlators into the products of the bilocal and threelocal ones. 
In Ref. [12], an ensemble of fields, whose irreducible correlators 
possess such 
properties, has been called ``minimally extended Gaussian ensemble''.

Our Eq. (7) means 
that {\it in the presence of the axionic term} the ensemble of 
fields in the London limit of AHM should also belong to this type. 
Thus, we conclude that when the axionic term is added, dual AHM 
in the London limit can really serve as a good model  
of gluodynamics in the confining phase. 
The results of the present Letter also 
shed some light to the structure of vacuum of AHM in the London 
limit in the presence of the axionic interaction.

\vspace{6mm}
{\large \bf Acknowledgments}
\vspace{3mm}

I appreciate discussions with many people, especially 
with Profs. H. Kleinert,   
M.I. Polikarpov, and Yu.A. Simonov. 
I would also like to thank the theory group of the 
Quantum Field Theory Department of the Institute of Physics of the 
Humboldt University of Berlin for kind hospitality and Graduiertenkolleg 
{\it Elementarteilchenphysik} for financial support.

\newpage
{\large \bf References}

\vspace{3mm}
\noindent
$[1]$~H. Kleinert, Phys. Lett. B 293 
(1992) 168; 
M.I. Polikarpov, U.-J. Wiese, and M.A. Zubkov, Phys. Lett. B 309 
(1993) 133; M. Kiometzis, H. Kleinert, and A.M.J. Schakel, 
Phys. Rev. Lett. 73 (1994) 1975; 
P. Orland, Nucl. Phys. B 428 (1994) 221;   
M. Kiometzis, H. Kleinert, and A.M.J. Schakel, 
Fortschr. Phys. 43 (1995) 697; 
H. Kleinert, Theory of 
Fluctuating Nonholonomic Fields and Applications: Statistical Mechanics 
of Vortices and Defects and New Physical Laws in Spaces with Curvature 
and Torsion, in: Proceedings of NATO Advanced Study Institute on 
Formation and Interactions of Topological Defects at the University 
of Cambridge, England, Plenum Press, New York, 1995 (preprint 
cond-mat/9503030 (1995));
M. Sato and S. Yahikozawa, Nucl. Phys. B 436 (1995) 100; 
M.N. Chernodub and M.I. Polikarpov, preprints hep-th/9710205 
and ITEP-TH-55-97 (1997).\\ 
$[2]$~E.T. Akhmedov, M.N. Chernodub, M.I. Polikarpov, and M.A. Zubkov, 
Phys. Rev. 
D 53 (1996) 2087.\\ 
$[3]$ E.T. Akhmedov, JETP Lett. 64 (1996) 82.\\  
$[4]$~D.V. Antonov, Mod. Phys. Lett. A 13 (1998) 659.\\ 
$[5]$~S. Mandelstam, Phys. Lett. 
B 53 (1975) 476; Phys. Rep. C 23 (1976) 245;   
G. 't Hooft, in: High Energy Physics, ed. A. Zichichi  
(Italy, Bologna, 1976); H. Kleinert, Gauge Fields in Condensed Matter, 
Vol. 1: Superflow and Vortex Lines. Disorder Fields, Phase Transitions 
(World Scientific Publishing Co., Singapore, 1989).\\
$[6]$~K. Lee, Phys. Rev. D 48 (1993) 2493.\\
$[7]$~H. Kleinert, Phys. Lett. B 246 (1990) 127.\\ 
$[8]$~J. Ho$\check {\rm s}$ek, 
Prague preprint REZ-TH-91/5, 
Phys. Rev. D 46 (1992) 3645, 
Czech. J. Phys. 43 (1993) 309.\\
$[9]$~A.M. Polyakov, Nucl. Phys. B 486 (1997) 23.\\
$[10]$C.G. Callan, Jr. and J.A. Harvey, Nucl. Phys. B 250 (1985) 427.\\
$[11]$D.V. Antonov, preprint hep-th/9804030 (1998).\\
$[12]$Yu.A. Simonov and V.I. Shevchenko, Phys. Atom. Nucl. 
60 (1997) 1201.\\
$[13]$H.G. Dosch, Phys. Lett. B 190 (1987) 177; H.G. Dosch and 
Yu.A. Simonov, Phys. Lett. B 205 (1988) 339; Yu.A. Simonov, 
Nucl. Phys. B 324 (1989) 67; H.G. Dosch, Prog. Part. Nucl. Phys. 
33 (1994) 121; Yu.A. Simonov, Phys. Usp. 39 (1996) 313.\\
$[14]$Yu.A. Simonov, Sov. J. Nucl. Phys. 50 (1989) 134.\\ 
$[15]$D.V. Antonov and Yu.A. Simonov, Int. J. Mod. Phys. A 11 (1996) 4401; 
D.V. Antonov, JETP Lett. 63 (1996) 398; 
Int. J. Mod. Phys. A 12 (1997) 2047; 
Phys. Atom. Nucl. 60 (1997) 299, 478.\\ 
$[16]$A. Di Giacomo and H. Panagopoulos, Phys. Lett. B 285    
(1992) 133; M. D'Elia, A. Di Giacomo, and E. Meggiolaro, Phys. Lett. 
B 408 (1997) 315; 
A. Di Giacomo, 
E. Meggiolaro, and H. Panagopoulos, Nucl. Phys. 
B 483 (1997) 371; Nucl. Phys. Proc. Suppl. A 54 (1997) 343.

\end{document}